\documentclass[]{elsart}
\usepackage{amssymb}
\usepackage{epsfig}

\begin{document}
\runauthor{Huffman \textit{et al.}}
\begin{frontmatter}

\title{Magnetically Stabilized Luminescent Excitations in Hexagonal 
	Boron Nitride}

\author[NIST]{P.\,R. Huffman\thanksref{Corr}}, 
\author[HARVARD]{C.\,R. Brome}, 
\author[HARVARD]{J.\,S. Butterworth}, 
\author[HARVARD]{S.\,N. Dzhosyuk},
\author[HMI]{R.~Golub}, 
\author[LANL]{S.\,K. Lamoreaux}, 
\author[HARVARD]{C.\,E.\,H. Mattoni}, 
\author[HARVARD]{D.\,N. McKinsey}, and
\author[HARVARD]{J.\,M. Doyle}

\address[NIST]{National Institute of Standards and Technology, 
	    Gaithersburg, MD 20899, USA}
\address[HARVARD]{Harvard University, Cambridge, MA 02138, USA}
\address[HMI]{Hahn-Meitner Institut, Berlin, Germany}
\address[LANL]{Los Alamos National Laboratory, Los Alamos, NM 87545, 
	    USA}
\thanks[Corr]{Correspondence and requests 
	for materials should be sent to PRH (paul.huffman@nist.gov).}
	    
\begin{abstract} 
Magnetically stabilized luminescence is observed in hexagonal boron
nitride.  The luminescence is induced by absorption of cold neutrons
and is in the visible region.  In the absence of a magnetic field, the
photon emission level is observed to decay over several hundred
seconds.  A fraction of this luminescence can be suppressed if the
temperature is $T \lesssim 0.6$~K and the magnetic field is $B \gtrsim
1.0$~T\@.  Subsequent to irradiation and suppression, luminescence can
be induced by an increase in $T$ or lowering of $B$.  Possible
explanations include stabilization of triplet states or the
localization and stabilization of excitons.
\end{abstract}

\end{frontmatter}

\section{Introduction}
The study of luminescence in hexagonal boron nitride (h-BN) informs us
about properties intrinsic to the material as well as its defects and
impurities.  Techniques used to study h-BN include electron
paramagnetic resonance\cite{gei64,kvk69}, thermally stimulated
luminescence\cite{lop92}, thermally stimulated conduction\cite{lop92},
and luminescence induced by light, cathodic, electric, ionic, and
radioactive sources\cite{lar56,lsg92,ksz75,Kob94}.  Identified defects
include F-centers, nitrogen vacancies and elemental impurities
\cite{ksz75,khu73,gkl89,gkl90,dzh69,emk81}.  Information about the
intrinsic nature of h-BN has been obtained through studies of the
near-band-gap luminescence\cite{tbs94}.

A continuum photon emission spectrum in h-BN has been observed as a
result of thermal neutron exposure which produces energetic He$^{+}$
(1.47~MeV) and $^{7}$Li$^{+}$ (0.84~MeV) ions in the material via the
$^{10}$B(n,$\alpha$)$^{7}$Li reaction.  Other irradiation studies used
1.0~MeV H$^{+}$ and 1.8~MeV He$^{+}$ ions and resulted in the
characterization of a continuum photon emission spectrum of
280~nm to 600~nm\cite{Kob94}.  Here we present observations of magnetic
suppression of cold neutron induced luminescence in h-BN at low
temperature.  The luminescence is stimulated by the by-products of
neutron capture and can be suppressed by application of a magnetic
field.  The studies we present here were prompted by the use of h-BN
as a neutron shielding material in an experiment to magnetically trap
ultracold neutrons (UCN)\cite{huf00}.

\section{Experimental Apparatus}
\begin{figure}[h]
\begin{center}
\epsfig{file=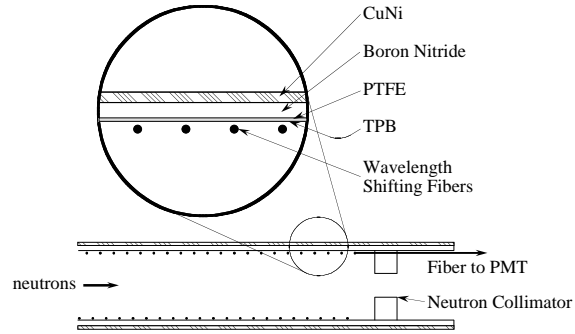,width=3.0in}
\end{center}
\caption{Neutron shielding and light collection system.  Neutrons 
enter from left, are collimated by a h-BN collimator, pass through the 
detection region and exit through a second collimator.  The detection 
system consists of an expanded PTFE diffuse reflector with a thin 
layer of TPB evaporated onto the inner surface.  The light is 
transported out of the apparatus using two helically coiled 
wavelength shifting fibers which are coupled into PMTs at room 
temperature.}
\label{fig:fibercell}
\end{figure}

Hexagonal boron nitride is used in our neutron trapping apparatus to
collimate an incident neutron beam and to absorb those neutrons
scattered out of the collimated beam.  Boron compounds are commonly
used as shielding materials for absorbing low energy neutrons. 
Natural boron has a 19.9~\% abundance of $^{10}$B, which has a large
thermal neutron absorption cross section, $\sigma_{\rm thermal} =
3840$~b.  Hexagonal boron nitride is particularly suitable because of
its high purity and mechanical properties.  Hexagonal BN can be
obtained with $<0.4~\%$ oxygen and $<0.1~\%$ all other total elemental
impurities\cite{combatbn}.  It is easily machinable and has a low
thermal expansion coefficient.  The h-BN used in our experiment is
obtained commercially as a solid cylinder and machined into the form
of interlocking tubular pieces with inner diameters of 3.86~cm and
outer diameters of 4.16~cm.  The assembled h-BN tube fits inside a
cupronickel (CuNi) cell body and extends its entire length, 110~cm
(see figure~\ref{fig:fibercell}).  Two cylindrical rings are used as
neutron collimators and are positioned at distances of 36~cm and 74~cm
relative to the front (beam entrance) of the h-BN tube.  These rings
each have an inner diameter of 2.0~cm, length of 1~cm and fit snugly
within the h-BN tube.  The cell is filled with liquid $^4$He and is
cooled by a dilution refrigerator to 0.1~K to 1.0~K\@.
 
A cold neutron beam (moderated to 40~K) traverses the cell parallel to
the long axis with a fluence of approximately $5 \times
10^{8}~{\mathrm s}^{-1}$.  About 60~\% of the neutrons entering the
cell are absorbed by the h-BN\@.  A set of superconducting magnets
resides just outside the cell body and when at maximum current, the
h-BN is in a magnetic field ranging from 1.0~T to 1.8~T\@.

Under normal operating conditions, a light collection system is
situated just inside the tube of h-BN between the two collimators (see
figure~\ref{fig:fibercell}).  This system consists of a diffuse
reflector made of expanded polytetrafluoroethylene (PTFE) onto which
is evaporated an ultraviolet-to-blue downconverter (tetraphenyl
butadiene, TPB)\cite{films} and two helically coiled blue-to-green
wavelength shifting fibers\cite{Kuraray}.  The fibers exit the
cryogenic apparatus and are each coupled to a photomultiplier tube. 
This system is sensitive to photons over the entire spectrum from the
extreme ultraviolet (EUV; $<100$~nm) to the visible blue ($\sim
500$~nm) and is not capable of further wavelength
discrimination\footnote{The fiber light collection system was not
actually used in the measurement made in Ref.~\cite{huf00}.}.

\section{Results}

In a typical run, the detection region is exposed to neutrons for
900~s.  During this irradiation period, the photomultiplier tubes
(PMTs) are turned off.  After 900~s, the neutron beam is blocked, the
PMTs are brought up to voltage and luminescence is detected.

\begin{figure}[h]
\begin{center}
\epsfig{file=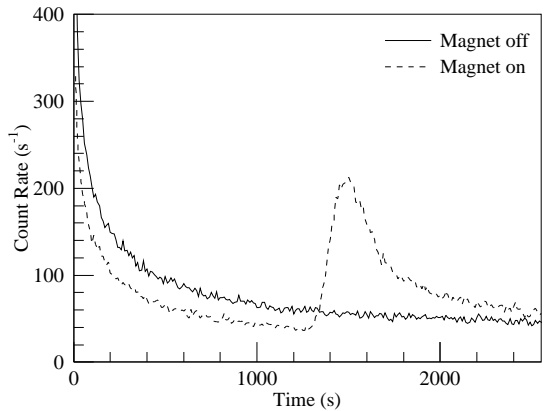,width=3.0in}
\end{center}
\caption{The time dependence of the luminescence signal with no
magnetic field (solid line) and when the magnetic field is energized
during the irradiation and deenergized 1275~s after the irradiation
ends (dashed line).}
\label{fig:nomagnet}
\end{figure}

The time dependence of the luminescence signal is shown in
figure~\ref{fig:nomagnet} (solid line), starting at the point when the
beam is turned off ($t=0$~s).  This decay does not fit to a single
exponential and varies roughly as inverse time.  When the magnet is
energized during the irradiation and observation, a significant
fraction of this luminescence signal is suppressed (see
figure~\ref{fig:nomagnet}, dashed line).  When the field is lowered
(starting at 1275~s), luminescence from the ``stored'' component is
emitted.

\begin{figure}
\begin{center}
\epsfig{file=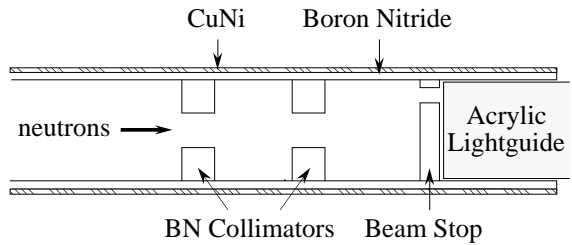,width=3.0in}
\end{center}
\caption{Schematic diagram of the simplified apparatus used to verify the 
luminescence was originating from the boron nitride.  The beam stop 
has three 4~mm slits machined at a radial distance of 14~mm from the 
center of the beam stop to allow the light to enter the acrylic 
lightguide.  No unscattered neutrons can reach the acrylic lightguide.}
\label{fig:cleanapp}
\end{figure}

\begin{figure}
\begin{center}
\epsfig{file=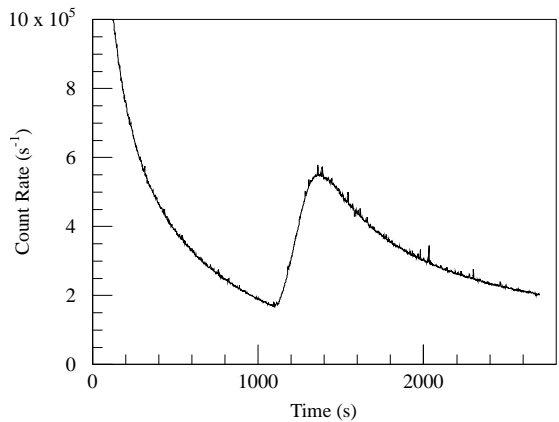,width=3.0in}
\end{center}
\caption{Luminescence signal using the setup shown in
figure~\ref{fig:cleanapp}.  The magnetic field is energized during the
irradiation and initial observation times and deenergized 1275~s after
the irradiation ends.}
\label{fig:cleandata}
\end{figure}

Numerous tests were performed to identify the source of the
luminescence signal and to characterize its behavior.  By the process
of elimination using multiple experimental setups, we determined that
the luminescence signal was originating from the boron nitride and
simplified the detection system to verify this conclusion.  In the
experimental setup shown in figure~\ref{fig:cleanapp}, only h-BN and
an acrylic lightguide were present (the TPB downconverter and
wavelength-shifting fibers were removed), and the geometry was
arranged such that neutrons could not directly impact the acrylic. 
The luminescence signal was observed.  Both components of this decay
signal can be seen in figure~\ref{fig:cleandata}.  In separate tests
performed at 4~K, we did not observe luminescence in an acrylic sample
under similar neutron irradiation conditions.

\begin{figure}
\begin{center}
\epsfig{file=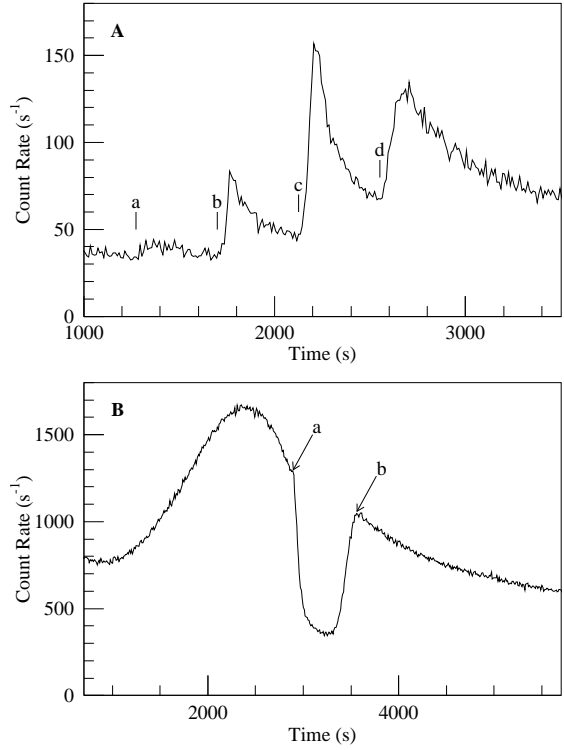,width=3.0in}
\end{center}
\caption{{\bf A}: The effect of lowering the magnetic field ($B$) in four 
steps as follows: a. $t=1275$~s, $B=100~\% \rightarrow 75~\%$;
b. $t=1700$~s, $B= 75~\% \rightarrow 50~\%$;
c. $t=2125$~s, $B= 50~\% \rightarrow 25~\%$;
d. $t=2550$~s, $B= 25~\% \rightarrow  0~\%$.  
{\bf B}: The magnetic field is slowly lowered
from its initial maximum value to zero over 2700~s. Just after this ramp 
ends, the field is rapidly raised to its maximum value beginning 
at point ``a'' and 570~s later, the field is lowered back to zero (point 
``b'').}
\label{fig:storage}
\end{figure}

Studies to characterize the behavior of the magnetically stabilized
component of the luminescence signal were performed using the fiber
cell apparatus (figure~\ref{fig:fibercell}).  Data resulting from
lowering the magnetic field in various ways is shown in
figure~\ref{fig:storage}.  The initial irradiation takes place with
the magnets at maximum current.  In figure~\ref{fig:storage}A the
magnet current is held constant for 800~s after the neutron beam is
turned off.  Then the magnetic field is reduced in four steps: from
$100~\% \rightarrow 75~\%$, $75~\% \rightarrow 50~\%$, $50~\% \rightarrow
25~\%$ and $25~\% \rightarrow 0~\%$ of its maximum value.  The largest
fraction of the luminescence is emitted in the last two reduction
steps.  Figure \ref{fig:storage}B further demonstrates the suppression
of luminescent emission.  In this measurement, the magnetic field is
slowly ramped down from 100~\% to zero over 2700~s.  Immediately after
this, the field is rapidly ($\sim 100$~s) turned back to 100~\% for
570~s and then back to zero.  As can be seen, the luminescence is
suppressed while the field is on.  We have verified, in a separate
experiment, that the magnetic field has no significant effect on the
efficiency of the detection system.

\begin{figure}
\begin{center}
\epsfig{file=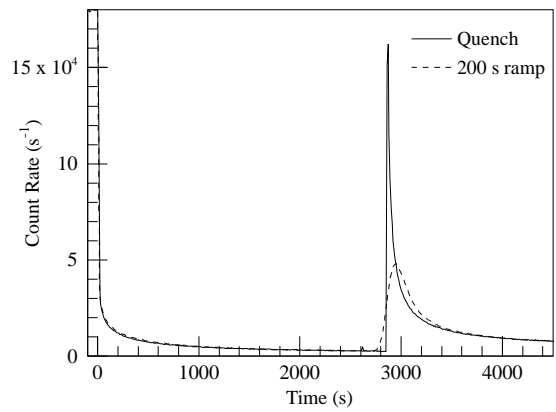,width=3.0in}
\end{center}
\caption{The time dependence
of the luminescence signal when the magnet is deenergized in less 
than one second (solid line) and the time dependence when the magnet 
is deenergized using a slower ramping speed of $\approx 200$~s (dashed 
line).}
\label{fig:quench}
\end{figure}

To characterize the time dependence of the decay of the stabilized
component, the superconducting magnet was quenched resulting in a
reduction in field from its maximum value to zero in less than 1~s. 
The time dependence of this decay signal is roughly the same as the
initial decay (see figure~\ref{fig:quench}).  In addition, the total
integrated number of counts using the magnet quench signal is roughly
equal to that obtained using a slow field ramp.  Presumably the
luminescence is the result of relaxation of some type of solid state
excitation in the h-BN\@.

The persistence of these excitations under conditions of high magnetic
field is most dramatically demonstrated by leaving the magnetic field
on for 28,800~s after irradiation.  After this time the magnetic field
is lowered to zero and luminescence is observed.  The number of
photons emitted after 28,800~s was within 50~\% of the number emitted
after only 1000~s of storage (under the same irradiation conditions).

\begin{figure}
\begin{center}
\epsfig{file=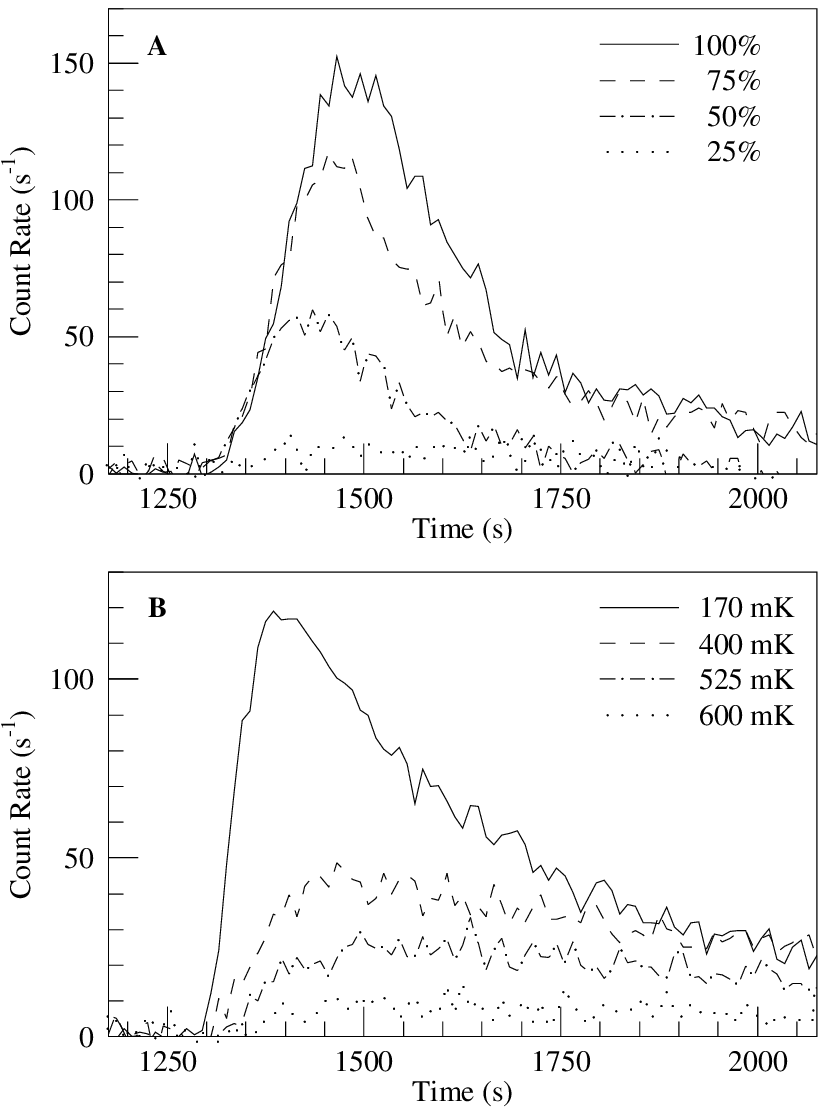,width=3.0in}
\end{center}
\caption{{\bf A}: The change in intensity of the luminescence signal
due to varying the magnitude of the magnetic field.  Each trace is for
a different initial magnetic field intensity.  In each case the magnet
is brought from its initial value to zero in 100~s.  {\bf B}: The
change in intensity of the luminescence signal due to varying the
temperature of the cell.  Each trace is for a different initial
temperature.  In each case the temperature is raised from its initial
value to 1~K while the magnet is kept at 100~\% of its full value.}
\label{fig:variation}
\end{figure}

We also observed that raising the temperature of the cell has a
similar effect to lowering the magnetic field.  Figure
\ref{fig:variation} shows data for different field and temperature
conditions.  In both sets of data the cell was irradiated by the
neutron beam for 900~s.  In figure \ref{fig:variation}A the magnetic
field is lowered from its initial value to zero in approximately 100~s
beginning 1275~s after the beam is turned off.  Traces are shown for
initial magnet currents of 100~\%, 75~\%, 50~\% and 25~\% of its maximum
current.  The data shown in figure \ref{fig:variation}B is taken
maintaining the magnet at its maximum current while raising the
temperature of the cell from its initial value to 1~K in approximately
100~s beginning 1275~s after the beam is turned off.  The traces are
for initial temperatures of 170~mK, 400~mK, 525~mK and 650~mK\@.  It is
important to note that temperature changes brought about by eddy
current heating while ramping the magnetic field ($< 20$~mK) are much
smaller than those required to initiate the release of luminescence
($> 200$~mK).

\section{Discussion}
Given the myriad of mechanisms for producing luminescence, we can only
offer general hypotheses about the magnetically stabilized component
of the luminescence signal and offer some general points.  The energy
of interaction of the magnetic moment of an electron or hole in h-BN
with a magnetic field of 1~T in temperature (Kelvin) units is 1.3~K\@. 
This is the same energy scale as that observed in
figure~\ref{fig:variation}.  Our highest applied field of 1.8~T
results in an energy of interaction $\sim 2.3$~K\@.  The applied
magnetic field should lead to polarization of paramagnetic species as
the thermal fluctuations present in the solid are not energetic enough
for depolarization.  In addition to electrons and holes, other
possibilities for paramagnetic species present in the h-BN include
triplet state excitons.  These typically have a magnetic moment of one
Bohr magneton ($1 \mu_{B}$) or greater.  Spin polarization of trapped
electrons, holes or excitons may take place and then affect the
possible relaxation paths to luminescence.

One possibility is that the magnetic field stabilizes a triplet state
of a defect center such as an F-center.  Triplet states can have very
long lifetimes\cite{tripst} and the stabilization would prevent
conversion to the much shorter lived singlet state.  Observations of
magnetic suppression of singlet state formation, resulting in enhanced
triplet state formation has been observed in alkali
halides\cite{Por71,Bal83}.  Unlike our case, these triplet states
exhibit pure exponential decay which is uneffected by magnetic field
changes.  Another possibility is that the interaction of an electron
or exciton with a paramagnetic center causes trapping, similar to that
seen in indium antimonide\cite{kst94}.  The applied magnetic field
could keep the two species aligned.  As the temperature is raised (or
the field lowered) thermal fluctuations would release the excitations
which could then undergo radiative recombination.  The apparent
distribution of trapping energies may indicate a combination of
thermoluminescent activity with a magnetic interaction.  That is, the
usual thermoluminescent trapping is modified by some type of magnetic
interaction, perhaps one of those discussed above.  In this case, the
spatially varying magnetic field would provide a continuum of trapping
energies.

\section{Conclusion}

We have observed a magnetically stabilized component in the
luminescence of hexagonal boron nitride under neutron irradiation. 
The excitations that give rise to the luminescence can be stabilized
for long periods by high magnetic fields ($\approx 1$~T) at low
temperatures ($\lesssim 600$~mK).  Further experiments are necessary
to elucidate the nature of the excitations, including spectroscopic
analysis of the luminescence, quantitative analysis of the magnetic
field dependence and optically detected electron paramagnetic
resonance.

\section{Acknowledgments}

This work was supported in part by the National Science Foundation
under grant No.  PHY-9424278.

\vspace{1.0in}

\vspace{0.5in}

\begin{thebibliography}{99}
\bibitem{combatbn}
AX05 grade h-BN, Carborundum Inc., Amherst NY. Certain trade names and
company products are mentioned in the text in order to adequately
specify the experimental procedure and equipment used.  In no case
does such identification imply recommendation or endorsement by NIST.
\end{thebibliography}
\end{document}